\documentclass[12pt]{article}
\usepackage{amssymb}
\parskip \baselineskip
\newcommand{\ds}{\displaystyle}

\newcommand{\F}{\mathbb{F}}

\newcommand{\ol}{\overline}

\newcommand{\ra}{\rightarrow}

\newcommand{\name}[2]{{#1}{\scriptsize{#2}}}

\begin{document}
\begin{center}
{\bf THE ASYMPTOTIC NUMBER OF BINARY CODES AND BINARY MATROIDS}

\name{M}{ARCEL} \name{W}{ILD}
\end{center}
\begin{quote}
A{\scriptsize BSTRACT}: {\footnotesize The asymptotic number of nonequivalent binary $n$-codes is determined. This is also the asymptotic number of nonisomorphic binary $n$-matroids.}
\end{quote}
{\bf 1. Introduction}

Recall that a {\it binary $n$}-code is a subspace $X$ of the $GF(2)$-vector space $V := GF(2)^n$. Two binary $n$-codes $X, X' \subseteq V$ are {\it equivalent} if for some permutation $\sigma$ of the symmetric group $S_n$ on $\{1, 2, \cdot \cdot, n\}$ we have
$$X' = X_\sigma := \{(x_{1 \sigma}, \cdots, x_{n \sigma} )| \ (x_1, \cdots, x_n) \in X\}$$
where $i\sigma$ is the image of $i$ under $\sigma$. Put
$$b(n) := \ \mbox{number of equivalence classes of binary} \ n\mbox{-codes}.$$
Notice that $b(n)$ also is the number of nonisomorphic binary matroids on a $n$-set [W]. The asymptotic behaviour of $b(n)$ was posed as open problem 14.5.4 in [O]. We chose the setting of codes to prove the main statement (1) below. 
For $K$ a field let
$$G(n,K):= \ \mbox{number of} \ K\mbox{-linear subspaces of} \ K^n.$$
We shall write $G(n, q)$ rather than $G(n, GF(q))$. Because each equivalence of binary $n$-codes has cardinality $\leq n!$ it follows that $b(n) \geq G(n,2)/n!$ for all $n$. It is claimed in [W, p.193] that for $n \ra \infty$ asymptotically

(1) \quad $b(n) \sim G(n,2)/n!$ .

For $\sigma \in S_n$ let $T_\sigma : V\ra V$ be the vector space automorphism defined on the canonical base by $T_\sigma (e_i) := e_{i\sigma}$. Let ${\cal L}(T_\sigma)$ be the lattice of all $T_\sigma$-invariant subspaces. Using the Cauchy-Frobenius lemma (falsely called Burnside's lemma) it follows that

(2) \quad $b(n)\quad =\quad \ds\frac{G(n,2)}{n!}  + \frac{1}{n!} \ds\sum_{\sigma \in S_n - \{id\}} | {\cal L} (T_\sigma)|$.

Hence proving (1) is equivalent to showing that

(3) \quad $\ds\sum_{\sigma \in S_n - \{id\}} |{\cal L}(T_\sigma)| \quad = \quad o (G(n,2))$.

There are ${n \choose 2}$ permutations $\sigma \in S_n$ with one $2$-cycle and $n-2$ many $1$-cycles. Any such $\sigma$ yields a $T_\sigma$ with at least $G(n-1,2)$ many invariant subspaces. Indeed, say $T_\sigma$ switches $e_1$ and $e_2$. Then the $n-1$ vectors $e_1 + e_2, \ e_3, \cdots, e_n$ are fixed by $T_\sigma$. Hence 

(4) \quad ${n \choose 2} G(n-1,2) \quad \mbox{is a {\it lower} bound for} \ \sum_{\sigma \in S_n - \{id\}} | {\cal L}(T_\sigma)|.$

This shows that (3) can only be true if $G(n,2)$ grows super-exponentially with $n$.
Proving (3) was undertaken in [W] but, as pointed out by Robert F. Lax [L], there is an error\footnote{Other than claimed, the $\tau := \sigma^{\mu_1}$ in the proof of (24) in [W] need not have $\rho (\sigma) \ 1$-cycles. Essentially, in the present article we replace $\rho (\sigma)$ by $n_1(\sigma)$. Then the proof of (24) goes through but claim (25) must be split into three subcases; see Lemma 4 below. Some minor shortcomings, one of them mentioned in [MR 1755766], are mended as well.} in the proof of [W, Lemma 6]. It is fixed in Section 2 below; not by introducing major new ideas but by partitioning $S_n$ more carefully. In section 3 statement (1) is refined so as to give the rate of convergence of $b(n)$ to $G(n, 2)/n!$. It has been claimed that (1) is an immediate consequence of [LPR] which establishes that almost all $n$-codes $X$ have a trivial automorphism group
$\mbox{Aut}(X) := \{ \sigma \in S_n| \ X_\sigma = X\}.$
Not so. The ``immediate''  implication rather goes in the opposite direction to some extent.  More details in Section 4.

\vskip 1cm

{\bf 2. The correct proof of $b(n) \sim G(n,2)/n!$}

Slightly modifying the proof of [W, Lemma 1] we have

{\bf Lemma 1:} For all prime powers $q$ there is a positive constant $C(q)$ such that
$$C(q) \cdot q^{n^2/4} \quad \leq \quad G(n, q) \quad \leq \quad 23\cdot q^{n^2/4} \quad \mbox{for all} \quad n \geq 0.$$
{\it Proof.} Let $q$ be fixed and put $G_n: = G(n,q)$. Note that $G_0 = 1, \ G_1 = 2$. According to [A, p.94] one has

(5) \quad $G_{n+1} = 2G_n + (q^n-1) G_{n-1} \quad (n\geq 1)$.

Letting $u_n := q^{-n^2/4}G_n \ (n \geq 0)$ it follows from (5) that
$$u_n \quad =  \quad 2q^{-n/2+1/4} u_{n-1} + (1-q^{-n+1})u_{n-2} \quad (n \geq 2).$$
Letting $\tau_n := 2q^{-n/2+1/4} +1-q^{-n+1}, \ a_n : = 2q^{-n/2+1/4} \tau_n^{-1}$, and $b_n := (1-q^{-n+1}) \tau_n^{-1} (n \geq 2)$, one deduces

(6) \quad $u_n \quad = \quad \tau_n (a_nu_{n-1} + b_n u_{n-2}) \quad (n \geq 2)$.

From $u_0 = 1, \ u_1 = 2q^{-1/4}, \ \tau_n > 1 \ (n\geq 2)$ and (6) follows $u_n \geq C(q) := \min \{u_0, u_1\}$\\
$(n \geq 0)$, whence the claimed lower bound in Lemma 1. Notice that $C(q) \ra 0$ for $q \ra \infty$.

As to the upper bound, from $u_0 = 1$ and $0 < u_1 = 2q^{-1/4} \leq 2 \cdot 2^{-1/4} < 1.7$ we get $a_2u_1 + b_2u_0 \in [0, 1.7]$, so (4) yields $u_2 \leq (1.7) \tau_2, \ u_3 \leq (1.7) \tau_2 \tau_3$ and so forth. One checks that $\tau_n = 2q^{-n/2+1/4}+1-q^{-n+1} \leq 2 \cdot 2^{-n/2+1/4} + 1-2^{-n+1}$ for all $n \geq 2$, whence $u_n \leq (1.7) \ds\prod_{n\geq 2} (2^{5/4-n/2} + 1-2^{-n+1})$ for all $n \geq 0$. By a brute force MATHEMATICA calculation the latter product is found to be smaller than $23$. \quad $\blacksquare$

Several people pointed out that lower and upper bounds for $G(n,q)$ can also be obtained by using (27). A more careful analysis [W] shows that 

(7) \quad $\ds\lim_{n \ra \infty} u_{2n} = 7.371969$ and $\ds\lim_{n \ra \infty} u_{2n+1} = 7.371949$ (rounded).

Both inequalities in Lemma 1 are important. The lower bound (with $q=2$) guarantees that our target (3) follows from Lemma 4 below; the upper bound will be used a lot along the way.

In order to get a hand on ${\cal L}(T_\sigma)$ we need the minimal polynomial
$$\min (T_\sigma, t) = \ds\prod_{i=1}^s p_i(t)^{\mu_i}$$
where the $p_i(t) \in GF(2)[t]$ are irreducible and $\mu_i \geq 1 \ (1 \leq i \leq s)$. Furthermore put
$$V_i := \ \mbox{ker} (p_i (T_\sigma)^{\mu_i}), \quad n_i := \dim (V_i) \quad (1 \leq i \leq s).$$
Then $V= V_1 \oplus \cdot \cdot \oplus V_s$ and from [W, p.196, 197] we have the following. If $T_i := (T_\sigma \upharpoonright V_i)$ then $T_i : V_i \ra V_i$ has minimal polynomial $\min (T_i, t) = p_i (t)^{\mu_i}$ and

(8) \quad ${\cal L}(T_\sigma) \quad \simeq \quad {\cal L}(T_1) \times  {\cal L}(T_2)  \ \ \times \cdot \cdot \times \ \ {\cal L}(T_s)$.

Assume that our $\sigma$ is a product of $r$ disjoint cycles $C_1, \cdots, C_r$ of lengths $2^{\alpha_j} \cdot u_j$ whereby $\alpha_j \geq 0$ and $u_j \geq 1$ is odd. If we standardize $p_1(t) := t+1$ then its corresponding parameters $\mu_1$ and $n_1$ satisfy

(9) \quad $\mu_1 = \max \{2^{\alpha_j}| \ 1\leq j \leq r\}$

and

(10) \quad $r\quad \leq \quad n_1 = 2^{\alpha_1} + 2^{\alpha_2} + \cdot \cdot + 2^{\alpha_r} \quad \leq \quad n$.

{\bf Lemma 2:} For all $\sigma \in S_n$ one has
$|{\cal L}(T_\sigma)| \leq | {\cal L}(T_1)| \cdot 2^{\frac{(n-n_1)^2}{8}+5n}.$

{\it Proof.} Since $T_i$ is bijective  we have $T_i^{\mu_i} \neq 0$, so $p_i (t) =t$ is impossible, so each $p_i(t)\ (2\leq i \leq s)$ has degree  $d_i \geq 2$. Fix $T_i : V_i \ra V_i$ with $2 \leq i \leq s$. According to [BF, Thm. 6] one can write $T_i = Q+S$ where $S: V_i \ra V_i$ is semisimple and $Q: V_i \ra V_i$ is nilpotent. Moreover, putting $K : = GF(2) [t]/p_i(t)$, the map $Q$ is $K$-linear in a natural sense and ${\cal L}(T_i) = {\cal L}_K (Q)$. Since $K \simeq GF(2^{d_i})$ and $\dim_K(V_i) = n_i/d_i$, it follows from Lemma 1 that
$$|{\cal L}(T_i)| \quad \leq \quad G\left( \frac{n_i}{d_i}, 2^{d_i} \right) \quad \leq  \quad 23 \cdot (2^{d_i})^{(n_i/d_i)^2/4} \quad = \quad 23 \cdot 2^{n_i^2/4d_i}.$$
Using (8) and $d_i \geq 2 \ (2 \leq i \leq s)$ we get
\begin{eqnarray*} |{\cal L}(T_\sigma)| & \leq & |{\cal L}(T_1)| (23\cdot 2^{n^2_2/8}) \cdots (23 \cdot 2^{n^2_s/8})\\
& \leq & |{\cal L}(T_1)| \cdot 23^n\cdot 2^{n^2_2/8+\cdot \cdot +n^2_s/8}\\
& \leq & |{\cal L}(T_1)| \cdot 2^{5n+(n_2+\cdots + n_s)^2/8}. 
\end{eqnarray*} \hfill $\blacksquare$

The trick to decompose $T_i$ as $S +Q$ with $Q$ nilpotent and ${\cal L}(T_i) = {\cal L}_K(Q)$ also works for $T_i = T_1$. In fact one verifies at once that $T_1 = I+(T_1+I)$ with $(T_1+I)^{\mu_1} = 0$ and ${\cal L}(T_1) = {\cal L}(T_1+I)$. However $d_i \geq 2$ is essential in the proof of Lemma 2; for $d_1 =1$ one only gets the triviality (in view of Lemma 1) $|{\cal L}(T_1)| = O(2^{n^2_1/4})$. Nonetheless, the upper bound
$$|{\cal L}(T_\sigma)|\quad =\quad O(2^{n_1^2/4}) \cdot O(2^{(n-n_1)^2/8 \ + \ 5n})$$
will turn out good enough for all $\sigma \in S_n$ with a ``small'' value $n_1(\sigma ) \leq n-6 \log n$ (see ${\cal D}_1$ in Lemma 4). For $\sigma$ with a bigger $n_1(\sigma)$ we shall need the better bounds for $|{\cal L}(T_1)|$ derived in  Lemma 3 and  distinguish three subcases according to the size of $r(\sigma)$ (see ${\cal D}_2, {\cal D}_3, {\cal D}_4$ in Lemma 4).

{\bf Lemma 3:} Let $\sigma \in S_n$ have $r$ disjoint cycles. With $T_1,  n_1, \mu_1$ derived from $T_\sigma$ as above, one has:
\begin{enumerate}
 \item [(a)] \quad $|{\cal L}(T_1)| \quad \leq \quad G(r, 2) \cdot G(n_1 -r, 2)$
\item[(b)] \quad $|{\cal L}(T_1)| \quad \leq \quad G(r, 2)^{\mu_1}$
\end{enumerate}
{\it Proof.} Let $W$ be any $K$-vector space with $\dim (W) = \ol{n}$, and $Q: W \ra W$ a linear nilpotent map, say $Q^{m-1} \neq Q^m = 0$. Let $Q_2: = Q \upharpoonright \ \mbox{im} \ (Q)$. Note that $Q_2 \neq Q^2$ but $\mbox{im} \ (Q_2) = \ \mbox{im} \ (Q^2)$. According to [BF, Thm.7] one has

(11) \quad ${\cal L}(Q) = \ds\bigcup_{X \in {\cal L}(Q_2)}[X, Q^{-1}(X)]$

where $Q^{-1}(X): = \{w \in W| \ Q(w) \in X\}$ and $[X, Q^{-1}(X)] := \{Y \in {\cal L}(W)| \ X \subseteq Y \subseteq Q^{-1} (X) \}$ is an interval of the lattice ${\cal L}(W)$ of all subspaces of $W$. Its length is

(12) \quad $\dim (Q^{-1}(X)) - \dim (X) \quad = \quad \dim (\ker Q) \quad = : \quad \kappa_1$.

Since $Q_2: \ \mbox{im} (Q) \ra \ \mbox{im}(Q)$ and $\dim (\mbox{im}Q)  = \ol{n} - \kappa_1$ it follows from (11) and (12) that

(13) \quad $|{\cal L}(Q)| \quad \leq \quad |{\cal L}(Q_2)| \cdot G(\kappa_1, K) \quad \leq \quad G(\ol{n} - \kappa_1, K) \cdot G(\kappa_1, K)$.

Iterating this idea observe that $\ker (Q_2) = \ker (Q) \cap \ \mbox{im}(Q)$, hence $\kappa_2 := \dim (\ker Q_2) \leq \kappa_1$.  Putting $Q_3 := Q_2 \upharpoonright \ \mbox{im} (Q_2)$ one deduces as above
$$|{\cal L}(Q_2) | \quad \leq \quad |{\cal L}(Q_3)| \cdot G(\kappa_2, K)$$
which, substituted into (13), yields
$$|{\cal L}(Q)| \quad \leq \quad |{\cal L}(Q_3)| \cdot G(\kappa_2, K) \cdot G(\kappa_1, K).$$
By induction and because of $|{\cal L}(Q_{m+1})| =1$ one gets
$$|{\cal L}(Q)| \quad \leq \quad G(\kappa_m, K) \cdots G(\kappa_2, K)\cdot G(\kappa_1, K)$$
where $\kappa_m \leq \kappa_{m-1} \leq \cdot \cdot \leq \kappa_2 \leq \kappa_1$ are defined in the obvious way. Therefore

(14) \quad $|{\cal L}(Q)| \quad \leq \quad G(\dim (\ker Q), K)^m$.

We are interested, for fixed $\sigma \in S_n$, in the case $K = GF(2), Q= T_1 +I,  \ W = V_1, \ \ol{n} = n_1, \ m = \mu_1$. To fix ideas suppose that $(2, 5, 7, 9)$ is one of the cycles of $\sigma$. It gives rise to exactly one nonzero $v \in V$ with $T_\sigma (v) = v$; namely $v:= e_2 +e_5 + e_7+e_9$. So $Q(v) = 0$. Thus clearly $\dim (\ker Q) = r$. See (7) for the relation between $r$ and $n_1$.    Claim (a) now follows from (13) in view of ${\cal L}(T_1) = {\cal L}(T_1 +I)$.  Claim (b) follows from (14). \hfill $\blacksquare$

Lemma 2 is an improvement of [W, p.199], and Lemma 3(a) respectively (b) has been mentioned without proof in [W, p.200, line 3] respectively [W, (36)]. 
Summarizing, Lemma 2 works exclusively for $T_i$ with $2 \leq i \leq s$ because then $d_i \geq 2$, and Lemma 3 works exclusively for $T_1$ because information about $\ker (Q)$ (where $T_i = S+Q$) is available only for $i =1$. Notice that $T_\sigma = T_1$ when $n_1=n$.  We mention that more than $\lfloor n/2\rfloor !2^n$ permutations $\sigma \in S_n$ have $n_1(\sigma) = n$. If $n$ happens to be a power of $2$ there are more than $(n-1)!$ of them. The next Lemma concludes the proof of (3) and hence of (1) (see introduction and Lemma 1).

{\bf Lemma 4:} $\ds\sum_{\sigma \in S_n - \{id\}} |{\cal L}(T_\sigma)| \quad = \quad o (2^{n^2/4})$. 

{\it Proof.} In the sequel $r=r(\sigma)$ and $n_1 = n_1 (\sigma)$. Putting

${\cal D}_1:= \{\sigma \in S_n \ |n_1 \leq n - 6 \log n\}$,

${\cal D}_2 := \{\sigma \in S_n \setminus {\cal D}_1 \ | 1 \leq r \leq 8 \log n_1\}$,

${\cal D}_3:= \{ \sigma \in S_n \setminus {\cal D}_1 \ | 8 \log n_1 < r < n_1- 8 \log n_1 \}$,

${\cal D}_4 := \{\sigma \in S_n \setminus {\cal D}_1 \ | n_1 - 8\log n_1 \leq r \leq n-1\}$

it suffices to verify the following:

(15) \quad $\ds\sum_{\sigma \in {\cal D}_1} |{\cal L}(T_\sigma)| = O(2^{(n^2/4) -n \log n})$,

(16) \quad $\ds\sum_{\sigma \in {\cal D}_2} | {\cal L}(T_\sigma ) | = O(2^{n^2/7})$,

(17) \quad $\ds\sum_{\sigma \in {\cal D}_3} | {\cal L}(T_\sigma)| = O (2^{(n^2/4)-n\log n})$,

(18) \quad $\ds\sum_{\sigma \in {\cal D}_4} |{\cal L}(T_\sigma)| = O(2^{(n^2/4)-(n/2) + 28\log^2n})$.

Without always mentioning it Lemma 1 will be used throughout the proof.
As to (15), because for fixed big enough $n$ the maximum of 
$$x \mapsto \frac{x^2}{4} + \frac{(n-x)^2}{8} + 5n \quad (0 \leq x \leq n-6\log n)$$ is obtained at $x = n-6 \log n$, it follows from Lemma 2 (and Lemma 1) that for all $\sigma \in {\cal D}_1$
$$|{\cal L}(T_\sigma)| \quad =\quad O(2^{\frac{n_1^2}{4} + \frac{(n-n_1)^2}{8}+5n})
\quad = \quad  O (2^{\frac{(n-6\log n)^2}{4} + \frac{(6 \log n)^2}{8}+5n})\quad = \quad O(2^{\frac{n^2}{4} - 2n \log n}), $$
which in view of $|{\cal D}_1| \leq n! \leq n^n  =  2^{n\log n}$ yields
$$\ds\sum_{\sigma \in {\cal D}_1} |{\cal L}(T_\sigma)|\quad = \quad 2^{n \log n} \cdot O(2^{\frac{n^2}{4}-2n\log n})\quad =\quad O(2^{\frac{n^2}{4}-n\log n}).$$
As to (16), from $r \leq 8 \log n_1 \leq 8 \log n$ and $\mu_1 \leq n$ (see (9)) and Lemma 3(b) one deduces
$$|{\cal L}(T_1)| \quad \leq \quad G(8\log n, 2)^n \ \leq \ \ds\left(23 \cdot 2^{(8 \log n)^2/4}\right)^n \quad = \ O(2^{16n\log^2 n\ + \ 5n}),$$
which by Lemma 2 gives
$$\ds\sum_{\sigma \in {\cal D}_2} |{\cal L}(T_\sigma)|\quad =\quad 2^{n \log n} \cdot  O(2^{16n\log^2n+ \frac{n^2}{8}+10n})\quad =\quad O(2^{\frac{n^2}{7}}).$$
As to (17), for all $\sigma \in {\cal D}_3$ one derives from Lemma 3(a) that
$$\begin{array}{lll}
|{\cal L}(T_1)| & \leq & G(r, 2) \cdot G(n_1-r, 2)\quad =\quad  O (2^{\frac{r^2}{4} + \frac{(n_1-r)^2}{4}}) \\
\\
& = & O(2^{\frac{(8\log n_1)^2}{4} + \frac{(n_1-8\log n_1)^2}{4}})\quad =\quad O (2^{\frac{n^2_1}{4} - 3 n_1 \log n_1}),\end{array}$$
so by Lemma 2
$$|{\cal L}(T_\sigma)|\quad =\quad O(2^{\frac{n^2_1}{4} - 3n_1 \log n_1 + \frac{(n-n_1)^2}{8}+5n})\quad =\quad O(2^{\frac{n^2}{4} - 3n_1 \log n_1+5n})\quad =\quad O(2^{\frac{n^2}{4} - 2n\log n}).$$
Here the last $=$ holds since $\sigma \in {\cal D}_3$ implies $\sigma \not\in {\cal D}_1$, so $n_1 > n-6 \log n$. As previously one now argues that
$$\ds\sum_{\sigma \in {\cal D}_3} |{\cal L}(T_\sigma)|\quad =\quad 2^{n \log n} \cdot O(2^{\frac{n^2}{4} - 2n \log n})\quad =\quad O(2^{\frac{n^2}{4} - n \log n}).$$
As to (18), for all $\sigma \in {\cal D}_4$ we use again $n_1 > n-6\log n$ and get

(19) \quad $r\quad \geq \quad n_1 - 8 \log n_1\quad > \quad (n -6\log n) - 8 \log n \quad = \quad n - 14 \log n$.

It follows that ${\cal D}_4$ is contained in the class ${\cal D}$ of all $\sigma \in S_n$ which have $\geq n -28 \log n$ many $1$-cycles. Indeed, if $\sigma \in {\cal D}_4$ had $< n - 28 \log n$ of them then $\sigma$ had $< (n - 28 \log n) + 14 \log n = n-14 \log n$ cycles altogether, contradicting (17). Thus
$$|{\cal D}_4| \quad \leq \quad |{\cal D}| \quad \leq \quad \ds{n \choose n-28\log n} [(28 \log n)!] \quad \leq \quad n^{28 \log n}.$$
Let $\sigma \in {\cal D}_4$. If $n_1 > n-12$ then by (8) and Lemma 3(a)
$$\begin{array}{lll} |{\cal L}(T_\sigma)| & \leq & |{\cal L}(T_1)| \cdot G(12,2) \quad \leq \quad G(n-1,2) \cdot G(1, 2)\cdot G(12,2)\\
\\ & = & O (2^{(n-1)^2/4})=O (2^{n^2/4\ - \ n/2}).
\end{array}$$
If $n_1 \leq n-12$ then by Lemma 2 
$$\begin{array}{lll} |{\cal L}(T_\sigma)| & \leq & |{\cal L} (T_1)| \cdot 2^{(n-n_1)^2/8\ + \ 5n}) \quad = \quad O(2^{n^2_1/4\ +\ (n-n_1)^2/8\ + \ 5n})\\
\\ & = & O(2^{(n-12)^2/4 \ + \ 5n})= O (2^{n^2/4 \ - \ n}). \end{array}$$
 It follows that

\hspace*{.5cm} $\ds\sum_{\sigma \in {\cal D}_4} |{\cal L}(T_\sigma)| \quad = \quad  |{\cal D}_4| \cdot O (2^{\frac{n^2}{4}-\frac{n}{2}}) \quad = \quad O(2^{28 \log^2n + \frac{n^2}{4}- \frac{n}{2}})$. \hfill $\blacksquare$

\vskip 1cm

{\bf 3. The main Theorem}

Let us strengthen (1) and state our main

{\bf Theorem:} For all sufficiently large $n$ one has
$$\ds\left( 1+2^{-\frac{n}{2} + 2 \log n + 1.2499}\right) \frac{G(n,2)}{n!}\quad \leq \quad b(n)\quad \leq \quad \left( 1+2^{-\frac{n}{2} + 2 \log n + 1.2501}\right) \frac{G(n,2)}{n!}.$$

{\it Proof:} The key ingredients are (7) and refinements of (4) and (18). As to enhancing (4), we need:

(20) \quad If $r(\sigma) = n-1$ then $|{\cal L}(T_\sigma)| = 2G(n-1,2) - G(n-2,2)$.

To see (20) consider w.l.o.g. the transposition $\sigma = (1,2)$. We claim that 

(21) \quad ${\cal L}(T_{(1,2)} ) = \{U \in {\cal L}(V) | \ \langle e_1 + e_2 \rangle \subseteq U$ or $U \subseteq \langle e_1 + e_2 \rangle^\perp \}$.

To see (21), let $U \in {\cal L}(T_{(1,2)})$ be such that $e_1 + e_2 \not\in U$. We have to show that $U \subseteq \langle e_1 + e_2 \rangle^\perp$. Assume to the contrary some $x = \ds\sum_{i=1}^n \lambda_i e_i$ in $U$ has scalar product $(e_1 + e_2)\cdot x \neq 0$. Then $x=e_1 + \ds\sum_{i=3}^n \lambda_i e_i$ or $x = e_2 + \ds\sum_{i=3}^n \lambda_ie_i$, say the former. From $T_{(1,2)} (x) = e_2 + \ds\sum_{i=3}^n \lambda_i e_i$ being in $U$ we get the contradiction $e_1 + e_2 = x + T_{(1,2)}(x) \in U$. This establishes $\subseteq$ in (21), and the similar $\supseteq$ is left to the reader.

By (21), ${\cal L}(T_{(1,2)})$ is the union of the $G(n-1,2)$-element interval sublattices $[\langle e_1 +e_2 \rangle, V]$ and $[0, \langle e_1 + e_2 \rangle^\perp ]$, whose intersection is the $G(n-2,2)$-element interval sublattice $[\langle e_1 +e_2 \rangle, \langle e_1 + e_2 \rangle^\perp]$. This gives (20). 

We are now in a position to double the lower bound in (4).
More precisely, because $G(n-2,2) = o(G(n-1, 2))$ it follows from (20) and (7) that

(22) \quad $\ds\sum_{r(\sigma) = n-1} |{\cal L}(T_\sigma)|\quad \geq\quad {n \choose 2} \cdot 2 \cdot 7.3719 \cdot 2^{\frac{(n-1)^2}{4}}$ \qquad ($n$ large).

By (2) this yields the lower bound in the Theorem; i.e. for large $n$ one has
$$b(n)\quad \geq \quad \ds\frac{G(n,2)}{n!} \left( 1+ {n \choose 2} 2^{-\frac{n}{2} + 1.25} \cdot \ds\frac{7.3719}{7.3720}\right)\quad \geq \quad \frac{G(n,2)}{n!} \ds\left( 1+ 2^{-\frac{n}{2} + 2 \log n + 1.2499}\right).$$
As to the upper bound, we turn around (22) (while increasing $7.3719$) and claim that

(23) \quad $\ds\sum_{r(\sigma) \leq n-1} |{\cal L}(T_\sigma)|\quad \leq \quad {n \choose 2} \cdot 2 \cdot 7.3720 \cdot 2^{\frac{(n-1)^2}{4}}$ \qquad ($n$ large).

By (2) inequality (23) implies indeed that for large $n$
$$b(n)\quad \leq \quad \frac{G(n,2)}{n!} \left( 1+ {n \choose 2} 2^{-\frac{n}{2} + 1.25} \cdot \frac{7.3720}{7.3719}\right)\quad \leq \quad \frac{G(n,2)}{n!} \ds\left( 1+ 2^{-\frac{n}{2} +2\log n + 1.2501}\right).$$
By (15), (16), (17) inequality (23) will be true if it holds with $\sigma$ just ranging over ${\cal D}_4$. Setting
$${\cal K} := \{ \sigma \in S_n | \ n_1 (\sigma) > n-6 \log n \ \mbox{and} \ n-14 \log n \leq r(\sigma) \leq n-1 \}$$
we have ${\cal D}_4 \subseteq {\cal K}$ by (19), and whence it suffices to show the following refinement of (18):
$$\ds\sum_{\sigma \in {\cal K}} | \ {\cal L}(T_\sigma)|\quad \leq \quad {n \choose 2} \cdot 2 \cdot 7.3720 \cdot 2^{\frac{n^2}{4} - \frac{n}{2} + \frac{1}{4}} \ (n \ \mbox{large}).$$
From (7) it is clear that $\geq$ in (22) switches to $\leq$ if $7.3719$ is replaced by $7.37197$. Hence it suffices to show that

(24) \quad $\ds\sum_{\sigma \in {\cal K}, \ r(\sigma) \leq n-2} | {\cal L}(T_\sigma)| = o(2^{\frac{n^2}{4} - \frac{n}{2}})$.

Let $\sigma \in S_n$ have $r(\sigma) = n-2$. Then it either is of type $\sigma = (1,2) (3,4)$ or $\sigma = (1, 2, 3)$. In the first case $T_\sigma = T_1$ and $|{\cal L}(T_\sigma)| \leq G(n-2, 2)\cdot G(2,2) = 5G(n-2,2)$ by Lemma 3(a). In the second case $n_1 = (n-3) +1 =n-2$ and $|{\cal L}(T_\sigma)| \leq 5G(n-2, 2)$ by (8). Hence 
$$\ds\sum_{r(\sigma) = n-2} \ |{\cal L}(T_\sigma)|\quad \leq \quad \left(3{n \choose 4} + 2{n \choose 3}\right) \cdot 5G(n-2,2)\quad =\quad o(2^{\frac{n^2}{4} - \frac{n}{2}}).$$
For $\sigma$ with $n-6 \leq r(\sigma) \leq n-3$ one verifies similarly that
$$\ds\sum_{n-6 \ \leq \ r(\sigma) \ \leq \ n-3} \ |{\cal L}(T_\sigma)|\quad =\quad O(n^{12} \cdot G(n-4,2))\quad =\quad o(2^{\frac{n^2}{4} - \frac{n}{2}}).$$
Now fix $\sigma \in {\cal K}$ with $r(\sigma) \leq n-7$. Consider $T_\sigma$ and the associated $T_1$. Putting $n_1:= n_1(\sigma)$ and $r:= r(\sigma)$, Lemma 3(a) implies
$$\begin{array}{lll} |{\cal L}(T_1)| & \leq & G(r,2) \cdot G(n_1 - r,2)\\
\\
& =& O(2^{\frac{r^2}{4} + \frac{(n_1-r)^2}{4}}) = O(2^{\frac{(n-7)^2}{4} + \frac{7^2}{4}}) = O(2^{\frac{n^2}{4} - (3.5)n}). \end{array}$$
Because $23(< 2^5)$ in Lemma 1 can be replaced by $7.372 (< 2^{2.9})$ it follows from Lemma 2 and $n_1 > n-6 \log n$ that
$$|{\cal L} (T_\sigma)|\quad \leq\quad 2^{\frac{(n-n_1)^2}{8}+ (2.9)n} \cdot O(2^{\frac{n^2}{4} - (3.5)n})\quad =\quad O(2^{\frac{n^2}{4} - (0.6)n+ \frac{36\log^2n}{8}}).$$
Like for $|{\cal D}_4|$ one argues that $|{\cal K}| \leq n^{28\log n}$, so
$$\ds\sum_{\sigma \in {\cal K}, \ r(\sigma) \leq n-7} |{\cal L}(T_\sigma)|\quad =\quad O(2^{\frac{n^2}{4} - (0.6)n + \frac{36\log^2n}{8} + 28 \log^2n})\quad =\quad o(2^{\frac{n^2}{4} - \frac{n}{2}}).$$
This proves (24) and whence the Theorem. \qquad $\blacksquare$

It should be clear from the proof that the exponents $1.2499$ and $1.2501$ in the Theorem {\it cannot} be replaced by $1.25 \pm \varepsilon$. However, equally clear, $1.25 \pm \varepsilon$ {\it can} be introduced if one distinguishes between even and odd integers. Thus for all $\varepsilon > 0$ and big enough $n$ one has

$$\begin{array}{lllll} \left( 1 + 2^{-n + 2 \log n + \frac{13}{4} - \varepsilon} \right) \ \ds\frac{G(2n,2)}{(2n)!} & \leq & b(2n) & \leq & \left( 1+2^{- n + 2 \log n + \frac{13}{4} + \varepsilon}\right) \ \ds\frac{G(2n,2)}{(2n)!}\\
\\
\mbox{and}\\
\\
\left( 1+2^{- n + 2 \log n + \frac{11}{4} - \varepsilon}\right) \ \ds\frac{G(2n+1,2)}{(2n+1)!} & \leq & b(2n+1) & \leq & \left( 1 + 2^{-n + 2 \log n + \frac{11}{4} + \varepsilon}\right) \ \ds\frac{G(2n+1,2)}{(2n+1)!} \end{array}$$

{\bf 4. About a result of Lefmann, Phelps and R\={o}dl}

Following [LPR] call the (binary) $n$-code $X$ {\it rigid} if Aut$(X) := \{ \sigma \in S_n : \ X_\sigma = X\}$ is trivial. 
Theorem 3 in [LPR] states the following:

(25) \quad Given $\alpha > 0$ there is $n_0$ such that for all fixed $n$ and $d$ with $n > n_0$ and \\
\hspace*{1.1cm} $(2+\varepsilon) \log n < d < n - (2+\varepsilon) \log n$ the proportion of  non-rigid $d$-dimensional   \\
\hspace*{1.1cm} $n$-codes is smaller than $\alpha$.

In other words, let $\alpha (n) := \max \{\alpha (n,d) : (2+\varepsilon) \log n < d < n-(2+\varepsilon) \log n \}$, where

(26) \quad $\alpha (n,d)\quad := \quad \ds\frac{\left| \left\{ X : \ X \ \mbox{is non-rigid} \ d\mbox{-dimensional} \ n\mbox{-code}\right\} \right|}{ G(n, 2, d )}$

and the so called {\it Gauss coefficient} $G(n, 2, d)$ is defined as the number of $d$-dimensional $n$-codes. It is e.g. proven in [LPR, p.115] that 

(27) \quad $2^{nd-d^2} \ \leq \ G(n,2,d) \ \leq \ 4 \cdot 2^{nd-d^2} \quad (1 \leq d \leq n)$

Clearly (25) amounts to say that $\alpha (n) \ra 0$ as $n \ra \infty$.
This is a nice result but, as promised in the introduction, let us argue that (25) does not imply (1) unless more about both the convergence of $\alpha (n)$ and the size of the nontrivial groups Aut$(X)$ is known. More specifically we show that (25) on its own allows the asymptotic value of $b(n)$ to be much bigger than $G(n,2)/n!$.

For the sake of notation, let us momentarily redefine $\alpha (n)$ as $\max \{\alpha (n,d) : 1 \leq d \leq n\}$, and still\footnote{This assumption can only decrease the number $b(n)$ of equivalence classes; we are busy showing that nevertheless $b(n)$ can be made huge. Notice that for $d \leq 2 \log (n)-2$ in fact $\frac{1}{2} - \varepsilon$ of all $d$-dimensional $n$-codes are {\it non-rigid}, see [LPR, Cor.1].} assume that $\alpha (n) \ra 0$ for $n \ra \infty$. The crucial point is that a priori $\alpha (n)$ may go to $0$ too slowly, say $\alpha (n) \geq 2^{-n}$ for almost all $n$. What's more, we don't know how big the nontrivial automorphism groups Aut$(X)$ tend to be. Many might have size close to $n!$. To simplify the calculation let's postulate all nontrivial Aut$(X)$ have size $\geq 2^{2n}$, which is minuscule compared to $n!$. (Still, more realistic would be a lower bound on the {\it average} size of the nontrivial Aut$(X)$.) Thus the equivalence class of any $n$-code $X$ has either cardinality $n!$ (if it is rigid) or cardinality $\leq n!2^{-2n}$ otherwise. It follows that for $n$ big enough

$\begin{array}{lll} (28) \qquad b(n) & \geq & \ds\frac{(1-\alpha (n)) G(n,2)}{n!} + \ds\frac{\alpha (n) G(n,2)}{n!2^{-2n}}\\
\\
& \geq & \ds\frac{2^{2n} \alpha (n) G(n,2)}{n!} \quad \geq \quad 2^n \cdot \frac{G(n,2)}{n!},  \end{array}$
 
whence certainly not $b(n) \sim G(n,2)/n!$.

Let us say a few words about the $2^{-n}$ above and how it relates to the proof of (25) in [LPR]. A moment's thought shows that $\alpha (n, d)$ is the fraction of $d$-dimensional $n$-codes $X$ admitting a permutation $\sigma \in S_n$ of prime order $p$ with $X_\sigma = X$. The proof of (25) distinguishes the main cases $p=2$ and $p> 2$. These are further split in several subcases (for instances according to the number $t$ of $p$-cycles in $\sigma$). Not surprising, the provable speed of $\alpha (n, d) \ra 0$ as $n$ goes to infinity depends on the size of $d$. From (27) one easily gets that, for each fixed $\varepsilon > 0$, 
$$G(n,2) \sim \sum \left\{G(n,2,d) : (\frac{1}{2} - \varepsilon ) n < d < (\frac{1}{2} + \varepsilon)n\right\}.$$
Thus, as one of $L, P, R$ has pointed out, rather than $\alpha (n)$ it suffices in (28) to consider $\alpha (n,d)$ with $d$ close to $\frac{n}{2}$. But that doesn't help; even for $\alpha (n, \frac{n}{2})$ there are three subcases where $L, P, R$ cannot do better than $\alpha (n, \frac{n}{2}) \leq 2^{-n}$ (see p.118 last line, and (42), (45) in [LPR]). So our hypothesis in (28) that $\alpha (n)$ be $\geq 2^{-n}$ did not come out of the blue. Note that nothing whatsoever about the sizes of the nontrivial Aut$(X)$ can be gleaned from [LPR].
 
Conversely our result (1) immediately yields a weaker version of [LPR, Thm.3] where $\alpha (n)$ is replaced by its {\it averaged} companion $\beta (n)$. Thus, let $\beta (n)$ be the fraction of the $G(n,2)$ many $n$-codes $X$ with $|\mbox{Aut}(X)| \geq 2$.
It follows that the total number $b(n)$ of equivalence classes satisfies

(29) \quad $b(n) \quad \geq \quad \ds\frac{\beta (n) G(n,2)}{n!/2} \  + \  \frac{(1-\beta (n)) G(n,2)}{n!} \quad = \quad  \frac{(1+\beta (n))G(n,2)}{n!}.$

By (1) this forces $\beta  (n) \ra 0$ as $n \ra \infty$. With a little extra effort even something about $\alpha (n)$ can be deduced, but we must narrow the range of $d$ that is given in (25). Namely, we claim:

(30) \quad Put $\alpha (n): = \max \{ \alpha (n, d) | \ \frac{n}{2} - (0.7) \sqrt{n} < d < \frac{n}{2} + (0.7) \sqrt{n} \}$\\
\hspace*{1.16cm} with $\alpha (n,d)$ as in (26). Then $\alpha (n) \ra 0$ as $n \ra \infty$.

To see (30) we deduce from (27) and Lemma 1 that
$$G(n, 2, \frac{n}{2} -(0.7) \sqrt{n}) \quad \geq \quad 2^{\frac{n^2}{4} -0.49n}\quad \geq \quad \gamma 2^{-0.49n} \cdot G(n,2) \quad (n \ \mbox{large}),$$
where $\gamma := 1/(7.4)$. Now deny (30). Then for some $\varepsilon > 0$ infinitely often $\alpha (n) \geq \varepsilon$. For such $n$ we have
$$b(n)\quad \geq\quad \ds\frac{\gamma 2^{-0.49n} \cdot G(n,2) \cdot \varepsilon}{n!/2}\quad +\quad \frac{(1-\varepsilon \gamma 2^{-0.49n}) G(n,2)}{n!}\quad =\quad \frac{(1+\varepsilon \gamma 2^{-0.49n})G(n,2)}{n!},$$
which contradicts the upper bound\footnote{In fact the upper bound $b(n) = \ds\left( 1+O \left( 2^{- \frac{n}{2} + 28 \log^2 n} \right) \right) \frac{G(n,2)}{n!}$, which follows at once from (15) to (18), suffices.} of $b(n)$ given in our Theorem.

Let $b(n, d)$ be defined as the number of equivalence classes of $n$-codes of dimension $d$. A similar argument establishes that for constant $c$ and large $n$ 

(31) \quad $\ds\frac{G(n, 2, d)}{n!} \quad \leq \quad b(n, d) \quad \leq \quad 2^{c^2+3} \cdot \frac{G(n, 2, d)}{n!} \ \  \left( \frac{n}{2} - c \leq d \leq \frac{n}{2} +c\right)$

The lower bound in (31) is clear. As to the upper bound, suppose that for some $d'$ in the range of (31) we have $b(n, d') \geq 2^{c^2+3} \cdot \ds\frac{G(n, 2, d')}{n!}$ for infinitely many $n$. Then by (27) on the one hand
$$b(n, d') \quad \geq \quad \ds\frac{2^{c^2+3} \cdot 2^{\frac{n^2}{4}-c^2}}{n!} \quad \geq \quad \frac{5\cdot 2^{\frac{n^2}{4}}}{n!},$$
while on the other hand
$$G(n, 2, d') \quad \leq \quad G(n, 2, \frac{n}{2}) \quad \leq \quad 4 \cdot 2^{\frac{n^2}{4}}.$$
In view of (7) this implies that for infinitely many $n$
$$\begin{array}{lll} b(n) & = & b(n, d') + \ds\sum_{d\neq d'} b(n, d)\ \ \geq \ \ \frac{5\cdot 2^{\frac{n^2}{4}}}{n!} \ + \ \frac{G(n,2) \  - \  G(n, 2, d')}{n!} \\
\\
& \geq & \ds\frac{5\cdot 2^{\frac{n^2}{4}}}{n!} \ + \ \frac{(7.37-4)2^{\frac{n^2}{4}}}{n!} \ \ = \ \ \ds\frac{(8.37)2^{\frac{n^2}{4}}}{n!}, \end{array}$$
which contradicts (1) and (7). One may replace $c^2+3$ by $c^2+2.0001$ in (31) but that's relevant at most when $c=0, \ d = \frac{n}{2}$ ($n$ even).

To summarize,  our proof of (1), developed in ignorance of [LPR], is shorter and somewhat neater than the proof of (25) in [LPR]. As shown in (28), there is no easy path from (25) to (1) (let alone from (25) to the Theorem in Section 3). But, witnessed by (29) and (30), an easy path leads from (1) close to (25).  The two approaches are  dual  in the following sense. The group $S_n$ operates on the set ${\cal L}(V)$. We focus on the set of fixpoints of $\sigma \in S_n$ (i.e. ${\cal L}(T_\sigma)$), whereas [LPR] focuses on the set of fixpoints of $X \in {\cal L}(V)$ (i.e. Aut$(X)$).

\begin{center}
{\bf References}
\end{center}
\begin{enumerate}
\item[{[A]}] M. Aigner,  Combinatorial  Theory, Springer 1979.
 \item [{[BF]}] L. Brickman and P.A. Fillmore, The invariant subspace lattice of a linear transformation, Canad. J. Math. {\bf 19} (1967) 810-822.
\item[{[L]}] R.F. Lax, On the character of $S_n$ acting on subspaces of $\F_q^n$, to appear in Finite Fields and their Applications.
\item[{[LPR]}] H. Lefmann, K.T. Phelps, V. R\"{o}dl, Rigid linear binary codes, Journal of Comb. Theory, Series A 63(1993) 110-128.
\item[{[O]}] J.G. Oxley, Matroid Theory, Oxford University Press 1997.
\item[{[W]}] M. Wild, The asymptotic number of inequivalent binary codes and nonisomorphic binary matroids, Finite Fields and their Applications {\bf 6} (2000) 192-202.
\end{enumerate}

\end{document}